\documentstyle[12pt]{article}
\def\slash#1{#1\kern -0.55em /}
\begin{document}
\large
\begin{flushright}{
}
\end{flushright}
\vfil
\vfil

\begin{center}
{\large{\bf A continuum limit of the chiral Jacobian in lattice gauge theory  }}
\end{center}
\vskip .5 truecm
\centerline{\bf Kazuo Fujikawa}
\vskip .4 truecm
\centerline {\it Department of Physics,University of Tokyo}
\centerline {\it Bunkyo-ku,Tokyo 113,Japan}
\vskip 0.5 truecm
\normalsize
\begin{abstract}
We study the implications of  the index theorem and chiral Jacobian  in lattice gauge theory , which have been formulated by Hasenfratz, Laliena and Niedermayer  and by  L\"{u}scher,  on the continuum formulation of the chiral Jacobian 
and anomaly. We take  a continuum limit of the lattice Jacobian factor without referring to  perturbative expansion and recover the  result of continuum theory 
by using only the general properties of the lattice Dirac operator.  
This procedure is based on a set of well-defined rules and thus provides an 
alternative approach to the conventional analysis of the chiral Jacobian and 
related anomaly in continuum theory. By using an explicit form of the lattice
Dirac operator introduced by Neuberger, which satisfies the Ginsparg-Wilson relation,  we illustrate our calculation  in some detail. We also briefly comment on the index theorem with a finite cut-off from the present viewpoint.  
\end{abstract}
\newpage
\section{Introduction}
There has been an interesting  development recently  in the treatment of fermions 
in  lattice gauge theory. This development is based on the lattice Dirac  
operator $D$ which satisfies the so-called Ginsparg-Wilson relation[1]
\begin{equation}
\gamma_{5}D + D\gamma_{5} = aD\gamma_{5}D
\end{equation}
where $\gamma_{5}$ is a hermitian chiral Dirac matrix. 
An explicit example of the operator satisfying (1) has been given by Neuberger[2].The operator has also been discussed as a fixed point  form of block transformations[3].
On the basis of this 
operator Hasenfratz, Laliena and Niedermayer[4] formulated an index theorem on the lattice with a finite cut-off.  Utilizing  this notion of the index on the lattice,  L\"{u}scher introduced a new kind of 
chiral transformation [5]
\begin{equation}
\delta\psi = i\alpha\gamma_{5}(1 -\frac{1}{2}aD)\psi, \ \ \ \delta\bar{\psi} = \bar{\psi}i\alpha (1 - \frac{1}{2}aD)\gamma_{5} 
\end{equation}
with an infinitesimal constant parameter $\alpha$. This transformation leaves the action in 
\begin{equation}
\int {\cal D}\bar{\psi}{\cal D}\psi \exp [- \sum a^{4} \bar{\psi}D\psi ]
\end{equation}
invariant due to the property (1), and  gives rise to the chiral Jacobian 
factor
\begin{equation}
J = \exp \{ -2i\alpha Tr \gamma_{5}(1 -\frac{1}{2}aD)\}
\end{equation}
The index theorem [4] shows that this Jacobian factor carries the correct chiral
anomaly.

In the meantime, Neuberger[6]  performed a general analysis of chiral 
anomaly in lattice gauge theory, which is a lattice counter part of the 
analysis on the basis of  families index in continuum theory [7].

In the present paper, we study the possible implications of the above development in lattice theory on the chiral Jacobian and anomaly in continuum 
theory[8]. 
We evaluate the index theorem and chiral anomaly in lattice 
theory without using the perturbative expansion. In the past analyses of 
the index theorem and chiral anomaly in lattice theory,  it has been customary to use the weak
coupling perturbation theory[1]: Namely , one first let the coupling constant $g$
small and then let the lattice spacing small. In this procedure, the characteristic features of  lattice theory such as the non-linear dispersion relation
for  Fermion spectrum  remain in Feynman amplitudes. In contrast, our 
present analysis examines the chiral Jacobian  factor in the limit $a \rightarrow 0$ with the coupling constant $g$ kept fixed.  We then recover the result of continuum theory by using only the general properties of the lattice Dirac operator. In this latter approach, the  field configurations of both of gauge fields and fermions contributing to the index theorem and chiral anomaly correspond to those of  continuum theory. We thus naturally recover the conventional treatment  of chiral Jacobian in continuum field theory on the basis of a set of well-defined rules.  We also briefly comment on the implications of  different field configurations contributing to
the index theorem in our approach and in Ref.[4] on the notion of the index theorem with  a finite cut-off. 

In the following, we first recapitulate  the essence of the continuum formulation and then compare it to a continuum limit of the lattice formulation.

\section{Index theorem and chiral anomaly in continuum theory}
We start with the QCD-type Euclidean path integral
\begin{equation}
\int {\cal D}\bar{\psi}{\cal D}\psi [{\cal D}A_{\mu}] \exp [ \int \bar{\psi}(i\slash{D}-m)\psi d^{4}x + S_{YM}]
\end{equation}
where $\gamma^{\mu}$ matrices are anti-hermitian with $\gamma^{\mu}\gamma^{\nu} +
\gamma^{\nu}\gamma^{\mu} = 2g^{\mu\nu} = - 2\delta^{\mu\nu}$, and $\gamma_{5}
= - \gamma^{1} \gamma^{2} \gamma^{3} \gamma^{4}$ is hermitian; $\slash{D}\equiv \gamma^{\mu}(\partial_{\mu}- igA^{a}_{\mu}T^{a}) =\gamma^{\mu}(\partial_{\mu}- igA_{\mu})$ with Yang-Mills generators $T^{a}$. $S_{YM}$ stands for the Yang-Mills action and $[{\cal D}A_{\mu}]$ contains a suitable gauge fixing.   

To analyze the chiral Jacobian we expand the fermion variables [8]
\begin{eqnarray}
\psi (x) &=& \sum_{n}a_{n}\varphi_{n}(x)\nonumber\\
\bar{\psi}(x) &=& \sum_{n}\bar{b}_{n}\varphi_{n}^{\dagger}(x)
\end{eqnarray}
in terms of the eigen-functions of hermitian $\slash{D}$
\begin{eqnarray}
\slash{D}\varphi_{n} (x) &=& \lambda_{n}\varphi_{n} (x)\nonumber\\
\int d^{4}x \varphi_{n}^{\dagger}(x)\varphi_{l} (x) &=& \delta_{n,l}
\end{eqnarray}
which diagonalize the fermionic action in (5).
The fermionic path integral measure is then written as 
\begin{equation}
{\cal D}\bar{\psi}{\cal D}\psi = \lim_{N\rightarrow \infty}\prod_{n=1}^{N}
d\bar{b}_{n}da_{n}
\end{equation}
Under an infinitesimal  global chiral transformation
\begin{equation}
\delta\psi = i\alpha\gamma_{5}\psi, \ \ \ \delta\bar{\psi} = \bar{\psi}
i\alpha\gamma_{5}
\end{equation}
we  obtain the Jacobian factor
\begin{eqnarray}
J &=& \exp [-2i\alpha  \lim_{N\rightarrow\infty}\sum_{n=1}^{N}\int d^{4}x \varphi_{n}^{\dagger}(x)\gamma_{5}\varphi_{n} (x)]\nonumber\\
&=& \exp [-2i\alpha (n_{+} - n_{-})]
\end{eqnarray}
where $n_{\pm}$ stand for the number of eigenfunctions with vanishing eigenvalues and  $\gamma_{5}\varphi_{n}= \pm \varphi_{n}$ in (5). We here used the relation $\int d^{4}x \varphi_{n}^{\dagger}(x)\gamma_{5}\varphi_{n} (x)=0$ for $\lambda_{n}\neq 0$. The Atiyah-Singer index theorem $n_{+} - n_{-}= \nu$ with  
 Pontryagin index $\nu$, which was  confirmed for one-
instanton sector in $R^{4}$ space by Jackiw and Rebbi[9], shows that the chiral
Jacobian contains the correct information of chiral anomaly. 

To extract a local version of the index (i.e., anomaly)[10], we start with the 
expression
\begin{eqnarray}
n_{+} - n_{-}&=& \lim_{N\rightarrow\infty}\sum_{n=1}^{N}\int d^{4}x \varphi_{n}^{\dagger}(x)\gamma_{5}f((\lambda_{n})^{2}/M^{2})\varphi_{n} (x)\nonumber\\
&=&\lim_{N\rightarrow\infty}\sum_{n=1}^{N}\int d^{4}x \varphi_{n}^{\dagger}(x)\gamma_{5}f(\slash{D}^{2}/M^{2})\varphi_{n} (x)\nonumber\\ 
&\equiv& Tr \gamma_{5}f(\slash{D}^{2}/M^{2})
\end{eqnarray}
for {\em any} smooth function $f(x)$ which rapidly goes to zero for $x=\infty$ with $f(0)=1$. Since $\gamma_{5}f(\slash{D}^{2}/M^{2})$ is a well-regularized operator, we may now use the plane wave basis of fermionic variables to extract an explicit gauge field dependence,  and we  define  a local version of the index  as
\begin{eqnarray}
&&\lim_{M\rightarrow\infty}tr \gamma_{5}f(\slash{D}^{2}/M^{2})\nonumber\\
&\equiv&\lim_{M\rightarrow\infty}\sum_{n=1}^{\infty} \varphi_{n}^{\dagger}(x)\gamma_{5}f(\slash{D}^{2}/M^{2})\varphi_{n} (x)\nonumber\\
&=& \lim_{M\rightarrow\infty}tr \int \frac{d^{4}k}{(2\pi)^{4}}e^{-ikx}\gamma_{5}f(\slash{D}^{2}/M^{2})e^{ikx}\nonumber\\
&=&\lim_{M\rightarrow\infty}tr \int \frac{d^{4}k}{(2\pi)^{4}}\gamma_{5}f\{
(ik_{\mu}+ D_{\mu})^{2}/M^{2} - \frac{ig}{4}[\gamma^{\mu}, \gamma^{\nu}]F_{\mu\nu}/M^{2}\}\nonumber\\
&=&\lim_{M\rightarrow\infty}tr M^{4}\int \frac{d^{4}k}{(2\pi)^{4}}\gamma_{5}f\{
(ik_{\mu}+ D_{\mu}/M)^{2} - \frac{ig}{4}[\gamma^{\mu}, \gamma^{\nu}]F_{\mu\nu}/M^{2}\} 
\end{eqnarray}
where the remaining trace stands for  Dirac and Yang-Mills indices. We also
used the relation
\begin{equation}
\slash{D}^{2} = D_{\mu}D^{\mu} - \frac{ig}{4}[\gamma^{\mu}, \gamma^{\nu}]F_{\mu\nu}
\end{equation}
and the rescaling of the variable $k_{\mu}\rightarrow M k_{\mu}$. 

By noting  $tr \gamma_{5}= tr \gamma_{5}[\gamma^{\mu}, \gamma^{\nu}]=0$, the above expression ( after expansion in powers of $1/M$) is written as ( with $ \epsilon^{1234}=1$)
\begin{eqnarray}
\lim_{M\rightarrow\infty}tr \gamma_{5}f(\slash{D}^{2}/M^{2})
&=&tr \gamma_{5}\frac{1}{2!}\{\frac{-ig}{4}[\gamma^{\mu}, \gamma^{\nu}]F_{\mu\nu}\}^{2} \int \frac{d^{4}k}{(2\pi)^{4}}f^{\prime\prime}(-k_{\mu}k^{\mu})\nonumber\\
&&= \frac{g^{2}}{32\pi^{2}}tr \epsilon^{\mu\nu\alpha\beta}F_{\mu\nu}F_{\alpha\beta}
\end{eqnarray}
where we used 
\begin{eqnarray}
\int \frac{d^{4}k}{(2\pi)^{4}}f^{\prime\prime}(-k_{\mu}k^{\mu})&=&
\frac{1}{16\pi^{2}}\int_{0}^{\infty} f^{\prime\prime}(x)xdx\nonumber\\
&=& \frac{1}{16\pi^{2}}
\end{eqnarray}
with $x= -k_{\mu}k^{\mu}>0$ in our metric. 
One can confirm that any finite interval $-L\leq k_{\mu} \leq L$ of momentum
variables in (12) {\em before} the rescaling $k_{\mu}\rightarrow M k_{\mu}$
gives rise to a vanishing contribution to (14). In this sense, the short 
distance contribution determines the anomaly.

When one combines (11) and (14), one establishes the Atiyah-Singer index theorem (in $R^{4}$ space). 
We  note that the local version of the index (anomaly)  is valid for Abelian theory also.
The global index (11) as well as a local version of the index (14) are both independent of the regulator  $f(x)$  provided [8] 
\begin{equation}
f(0) =1, \ \ \ f(\infty)=0,\ \ \ f^{\prime}(x)x|_{x=0}=f^{\prime}(x)x|_{x=\infty}=0 
\end{equation}
Our regulator $f(x)$ imposes gauge invariance, and thus the regulator independence  of chiral anomaly  is consitent with the analysis of Adler[11], who showed that the chiral anomaly is independent of divergence  and perfectly finite and  well-defined if one imposes gauge invariance on the triangle diagram. 
 
Another way to see the role of $f(x)$ is to understand it as a regulator of 
Noether currents. This was illustrated  in connection with the generalized Pauli-
Villars regularization[12][13] elsewhere[14].

\section{Index theorem on the lattice}
We assume the existence of a lattice Dirac operator satisfying the relation
\begin{equation}
\gamma_{5}D + D\gamma_{5} = aD\gamma_{5}D
\end{equation}
with hermiticity property , $(\gamma_{5}D)^{\dagger}= D^{\dagger}\gamma_{5}=
\gamma_{5}D$.
One can then confirm the relation (for half of the chiral Jacobian factor 
in (4))[4]
\begin{eqnarray}
Tr [ \gamma_{5}(1-\frac{1}{2}aD)]&=& \sum_{n}\{\phi^{\dagger}_{n}\gamma_{5}\phi_{n} - \frac{1}{2}\phi^{\dagger}_{n}\gamma_{5}aD\phi_{n}\}\nonumber\\
&=& \sum_{ \lambda_{n}=0}\phi^{\dagger}_{n}\gamma_{5}\phi_{n} + \sum_{ \lambda_{n}\neq 0}\phi^{\dagger}_{n}\gamma_{5}\phi_{n} - \sum_{n}\frac{1}{2}a\lambda_{n}\phi^{\dagger}_{n}\phi_{n}\nonumber\\
&=&\sum_{ \lambda_{n}=0}\phi^{\dagger}_{n}\gamma_{5}\phi_{n}\nonumber\\
&=& n_{+} - n_{-} =  index
\end{eqnarray}
where $n_{\pm}$ stand for the number of  normalizable zero modes in 
\begin{equation}
\gamma_{5}D\phi_{n}=\lambda_{n}\phi_{n}
\end{equation}
for the {\em hermitian}  operator $\gamma_{5}D$ with simultaneous eigenvalues $\gamma_{5}\phi_{n}= \pm \phi_{n}$.  
We also used the relation 
\begin{equation}
\phi^{\dagger}_{n}\gamma_{5}\phi_{n} = \frac{a}{2}\lambda_{n}\phi^{\dagger}_{n}\phi_{n}
\end{equation}
for $\lambda_{n}\neq 0$, which is derived by sandwiching the relation(17) by 
$\phi^{\dagger}_{n}$ and $\phi_{n}$.  
Here the inner product $\phi^{\dagger}_{n}\phi_{n} = (\phi_{n},\phi_{n})
\equiv \sum_{x} a^{4}\phi^{\star}_{n}(x)\phi_{n}(x)$ is defined by summing over all the lattice points, which are not explicitly written
in $\phi_{n}$. 
Note that one has $\gamma_{5}D(\gamma_{5}\phi_{n})=0$ if $\gamma_{5}D\phi_{n}=0$ by using the relation (17), namely
$\gamma_{5}D[(1\pm\gamma_{5})\phi_{n}]=0$. We can thus choose the eigenvectors
of $\gamma_{5}D$ with vanishing eigenvalues to be the eigenstates of $\gamma_{5}$, i.e., $(1\pm \gamma_{5})/2 \phi_{n}$.  To define the zero modes precisely, we assume that the lattice size is finite,  $L=Na= finite$.

The notion of index means that we transfer the information carried by gauge fields to fermionic variables, and it is sometimes more convenient to retain  $Tr \gamma_{5}$ as in (18) when evaluating the Jacobian factor. Also, the fermion measure of  path integral generally becomes non-trivial in the analysis of  index even in lattice theory.  

Next we investigate a local version of the index theorem to show that the index (18) is related to Pontragin index. We  start with
\begin{equation}
Tr\{\gamma_{5}[1-\frac{1}{2}aD]f(\frac{(\gamma_{5}D)^{2}}{M^{2}})\}
=n_{+} - n_{-}
\end{equation}
Namely, the index is not modified by any  regulator $f(x)$ with $f(0)=1$, as can be confirmed by using the basis in (19). 
The operator $\gamma_{5}D$ plays a priviledged role in the present analysis
of the index theorem.
We then consider a local version of the index
\begin{equation}
tr\{\gamma_{5}[1-\frac{1}{2}aD]f(\frac{(\gamma_{5}D)^{2}}{M^{2}})\}
\end{equation}
where trace stands for Dirac and Yang-Mills indices. 
A local version of the index is not sensitive to the precise boundary condition , and  one may take the infinite volume limit $L=Na \rightarrow\infty$ in the above expression. 

We now examine the continuum limit $a\rightarrow 0$ of the above local expression (22)\footnote{This continuum limit corresponds to the so-called ``naive'' continuum limit in the context of lattice gauge theory.}. We first observe that the term
\begin{equation}
tr\{\frac{1}{2}a\gamma_{5}Df(\frac{(\gamma_{5}D)^{2}}{M^{2}})\}
\end{equation}
goes to zero in this limit. The large eigenvalues of $\gamma_{5}D$ are 
truncated at the value $\sim M$ by the regulator $f(x)$ which rapidly goes to zero for large $x$. In other words, the global index of the operator $Tr \frac{a}{2}\gamma_{5}Df(\frac{(\gamma_{5}D)^{2}}{M^{2}})\sim O(aM)$.

We thus examine the small $a$ limit of 
\begin{equation}
tr\{\gamma_{5}f(\frac{(\gamma_{5}D)^{2}}{M^{2}})\}
\end{equation}
The operator appearing in this expression is well regularized by the function
$f(x)$ , and  we evaluate the above trace by using the plane wave basis to extract an explicit gauge field dependence.
We consider a square lattice where the momentum is defined in the Brillouin zone
\begin{equation}
-\frac{\pi}{2a}\leq k_{\mu} < \frac{3\pi}{2a}
\end{equation}
We assume that the operator $D$ is free of  species doubling; in other words, the operator $D$ blows up rapidly ($\sim \frac{1}{a}$) for small $a$ in the momentum region corresponding to species doublers. The contributions of doublers are  
eliminated by the regulator $f(x)$ in the above expression. We thus examine the above trace in the momentum range of the physical species
\begin{equation}
-\frac{\pi}{2a}\leq k_{\mu} < \frac{\pi}{2a}
\end{equation}

We now obtain the limiting $a\rightarrow 0$ expression
\begin{eqnarray}
&&\lim_{a\rightarrow 0}tr\{\gamma_{5}f(\frac{(\gamma_{5}D)^{2}}{M^{2}})\}\nonumber\\
&=& \lim_{a\rightarrow 0}tr \int_{-\frac{\pi}{2a}}^{\frac{\pi}{2a}}\frac{d^{4}k}{(2\pi)^{4}}e^{-ikx}\gamma_{5}f(\frac{(\gamma_{5}D)^{2}}{M^{2}})e^{ikx}\nonumber\\
&=&\lim_{L\rightarrow\infty}\lim_{a\rightarrow 0}tr \int_{-L}^{L}\frac{d^{4}k}{(2\pi)^{4}}e^{-ikx}\gamma_{5}f(\frac{(\gamma_{5}D)^{2}}{M^{2}})e^{ikx}\nonumber\\
&=&\lim_{L\rightarrow\infty}tr \int_{-L}^{L}\frac{d^{4}k}{(2\pi)^{4}}e^{-ikx}\gamma_{5}f(\frac{(-i\gamma_{5}\slash{D})^{2}}{M^{2}})e^{ikx}\nonumber\\
&=&tr\{\gamma_{5}f(\frac{\slash{D}^{2}}{M^{2}})\}
\end{eqnarray}
where  we first take the limit $a\rightarrow 0$ with fixed $k_{\mu}$ in 
$-L\leq k_{\mu} \leq L$, and then take the limit $L\rightarrow \infty$. This 
procedure is justified if the integral is well convergent\footnote{
To be precise, we deal with an integral of the structure 
$\int^{\frac{\pi}{2a}}_{-\frac{\pi}{2a}}dx f_{a}(x) =\int^{\frac{\pi}{2a}}_{L}dx f_{a}(x) + \int^{L}_{-L}dx f_{a}(x) + \int^{-L}_{-\frac{\pi}{2a}}dx f_{a}(x)
$
where $f_{a}(x)$ depends on the parameter $a$. ( A generalization to a 4-dimensional integral is straightforward.) We thus have to prove that both of 
$\lim_{a\rightarrow 0}\int^{\frac{\pi}{2a}}_{L}dx f_{a}(x)$ and $\lim_{a\rightarrow 0}\int^{-L}_{-\frac{\pi}{2a}}dx f_{a}(x)$ can be made arbitrarily small
if one let $L$ large. 
A typical integral we encounter in lattice theory has a generic structure
 $\lim_{a\rightarrow 0}
\int_{-\pi/2a}^{\pi/2a} dx e^{-\sin^{2}ax/(a^{2}M^{2})} = \lim_{L\rightarrow \infty}\int_{-L}^{L} dx e^{-x^{2}/M^{2}}$ 
and satisfies the above criterion, if one chooses the regulator $f(x)=e^{-x}$ . For  species doublers, $\lim_{a\rightarrow 0}
\int_{\pi/2a}^{3\pi/2a} dx e^{-1/(a^{2}M^{2})} = \lim_{a\rightarrow 0}
(\pi/a)  e^{-1/(a^{2}M^{2})} =0 $ due to  our assumed property of $D$.}
. We also assumed that the operator $D$ satisfies  the following relation in the limit $a\rightarrow 0$
\begin{eqnarray}
De^{ikx}g(x) &\rightarrow& e^{ikx}(\slash{k} -i\slash{\partial} -\slash{A})g(x)\nonumber\\
&=& -i\slash{D}(e^{ikx}g(x))
\end{eqnarray}
for any  {\em fixed} $k_{\mu}$, ($-\frac{\pi}{2a}< k_{\mu}< \frac{\pi}{2a}$), and a sufficiently smooth function $g(x)$. The function $g(x)$ corresponds to the gauge potential in our case, which in turn means that the gauge potential $A_{\mu}(x)$
is assumed to   vary very little  over the distances of the elementary lattice spacing. It is shown in eq.(62) of  Section 4 that an explicit example of $D$ 
given by Neuberger satisfies the property (28) without species doublers.

Our final expression (27) in the limit $M\rightarrow\infty$ combined with (21) thus reproduces the index theorem in the continuum formulation, (11) and (14), 
by using  the  quite general properties of the basic operator $D$ only: The basic relation (1) with hermitian $\gamma_{5}D$ and the continuum limit property (28) 
{\em without} species doubling in the limit $a\rightarrow 0$.
This shows  that  the Jacobian factor (18) in fact contains the correct chiral
anomaly[5]. (We are  implicitly assuming that the index (18) does not change in
the process of taking a continuum limit.)   

We now add several  comments on the present analysis. As is explained in the next section, the relation $Tr\gamma_{5} =0$  carries important information on species doubling if one uses suitable basis vectors to define it.  This relation is of course expected to hold for any 
basis in  finite theory, and we obtain by using the basis in (19) and the relation (20)
\begin{eqnarray}
Tr\gamma_{5} &=& \sum_{n} \phi^{\dagger}_{n}\gamma_{5}\phi_{n}\nonumber\\
&=& \sum_{ \lambda_{n}=0}\phi^{\dagger}_{n}\gamma_{5}\phi_{n} +
\sum_{ \lambda_{n}\neq 0}\phi^{\dagger}_{n}\gamma_{5}\phi_{n}\nonumber\\
&=& \sum_{ \lambda_{n}=0}\phi^{\dagger}_{n}\gamma_{5}\phi_{n}
+ \sum_{ \lambda_{n}\neq 0}\frac{a}{2}\lambda_{n}\nonumber\\
&=& n_{+} - n_{-} +  \sum_{ \lambda_{n}\neq 0}\frac{a}{2}\lambda_{n}=0
\end{eqnarray}
Namely,
\begin{equation}
n_{+} - n_{-} = -  \sum_{ \lambda_{n}\neq 0}\frac{a}{2}\lambda_{n} 
\end{equation}
This relation (30) shows that the chirality asymmetry $n_{+} - n_{-}$ at low energies is balanced by the chirality asymmetry at short distances $\sim a$ : In the limit $a \rightarrow 0$, only the eigenvalues of the order of $1/a$ contribute to the right-hand side of (30). 
This analysis is consistent with the analysis in  continuum[8]; we showed that only the short distances contribute to the calculation in (12). 
Incidentally, we note that 
\begin{equation}
Tr \gamma_{5}f((\gamma_{5}D)^{2}/M^{2}) = n_{+}- n_{-} +
\sum_{ \lambda_{n}\neq 0}\frac{a}{2}\lambda_{n}f(\lambda_{n}^{2}/M^{2})\neq 0
\end{equation}
in general.

It is instructive to consider an operator $D$ which satisfies the  relation
\begin{equation}
\gamma_{5}D + D\gamma_{5}=aD\gamma_{5}Dg(\gamma_{5}D)
\end{equation}
with  $g(x), \ \ g(0)\neq 0$,  which rapidly goes to zero at large $x$; namely, $g(x)$ contains a fixed mass scale other than $1/a$ and rapidly goes to zero for $x$ much larger than the fixed mass scale.  We still obtain an index relation
\begin{equation}
n_{+} - n_{-}=Tr\{\gamma_{5}[1-\frac{1}{2}aDg(\gamma_{5}D)]\}
\end{equation}
but by repeating the analysis in (29) , we have $n_{+} - n_{-}= 0$;
\begin{equation}
n_{+} - n_{-}= - \sum_{ \lambda_{n}\neq 0}\frac{1}{2}a\lambda_{n}g(\lambda_{n})\rightarrow 0
\end{equation}
for $a\rightarrow 0$. In this case, the chiral symmetry breaking in the operator $D$ is ``soft''
and all the species doublers survive at short distances ( or large eigenvalues) to lead to  vanishing chiral anomaly.

Although we here followed the standard procedure of deriving a  local version of index theorem in  continuum [10] as in eq.(21), one may of course expect that an explicit evaluation of the lattice Jacobian factor gives rise to the same result. We illustrate this calculation  in the next section by using an explicit formula of $D$ introduced by Neuberger. It is seen there that the basic relation (1) by itself does not specify the coefficient of  chiral anomaly uniquely.

\section{Chiral anomaly on the lattice}
The operator $D$ introduced by Neuberger[2] , which satisfies the relation (1), has an explicit expression
\begin{equation}
aD= 1 - \gamma_{5}\frac{H}{\sqrt{H^{2}}} =1 + X\frac{1}{\sqrt{X^{\dagger}X}}
\end{equation}
where $X=-\gamma_{5} H$ is the Wilson operator for a massive fermion
\begin{eqnarray}
X_{nm}&=&i\gamma^{\mu}C_{\mu}(n,m) + B(n,m) -\frac{1}{a}m_{0}\delta_{n,m}\nonumber\\
C_{\mu}(n,m)&=&\frac{1}{2a}[\delta_{m+\mu,n}U_{\mu}(m) - \delta_{m,n+\mu}U^{\dagger}_{\mu}(n)]\nonumber\\
B(n,m)&=&\frac{r}{2a}\sum_{\mu}[2\delta_{n,m}-\delta_{m+\mu,n}U_{\mu}(m)
-\delta_{m,n+\mu}U^{\dagger}_{\mu}(n)]\nonumber\\
U_{\mu}(m)&=& \exp [iagA_{\mu}(m)]
\end{eqnarray}
The chiral Jacobian is given by 
\begin{equation}
\ln J = -2i\alpha Tr[\gamma_{5}(1-\frac{a}{2}D)]
\end{equation}
In the lattice theory with a finite size $L=Na=finite$, one may expect $Tr\gamma_{5}=0$. Nevertheless, it is instructive to define this operation more precisely. For this purpose, we first observe an (analytic) index theorem\footnote{
A proof of this theorem goes as follows:  For non-vanishing eigenvalues in $M^{\dagger}Mu_{n}= \lambda_{n}^{2}u_{n}$, we have $MM^{\dagger}(Mu_{n}/\lambda_{n}) =  \lambda_{n}^{2}(Mu_{n}/\lambda_{n})\equiv \lambda_{n}^{2}v_{n}$. Namely, the finite dimensional (hermitian) square matrices $MM^{\dagger}$ and $M^{\dagger}M$ have the same number of non-vanishing eigenvalues, and consequently the same number of eigenfunctions with vanishing eigenvalues. The relation $(u_{n}, M^{\dagger}Mu_{n}) = (Mu_{n}, Mu_{n})$ shows that  $M^{\dagger}Mu_{n}=0$ implies  $Mu_{n}=0$, and the converse is obvious; similarly for $MM^{\dagger}v_{n}=0$ and $M^{\dagger}v_{n}=0$. We thus obtain the theorem.}
\begin{equation}
index\ \ of\ \ M = dim\ ker\ M - dim\ ker\ M^{\dagger}=0
\end{equation}
for any {\em finite} dimensional square matrix $M$. Here $dim\ ker M$, for example,
stands for the number of normalizable vectors with $Mu_{n}=0$.  This index theorem is 
responsible for the inevitable appearance of an infinite number of fields in the generalized Pauli-Villars regularization of a chiral fermion [12][13]: The operator $M$ corresponds to the fermion mass matrix there, and the above index corresponds to the number of chiral fermions. 

In our application, we consider a finite dimensional square matrix
\begin{equation}
M = \gamma^{\mu}C_{\mu}(n,m)(\frac{1+\gamma_{5}}{2})
\end{equation}
which satisfies $M^{\dagger} = (\frac{1+\gamma_{5}}{2})\slash{C}= \slash{C}(\frac{1- \gamma_{5}}{2})$ by noting $\gamma_{5}\slash{C}+\slash{C}\gamma_{5}=0$ for  $C_{\mu}$ in (36).
Note that $\slash{C}$ projects  fermion fields on a  lattice to the same
set  of fermion fields  on the same lattice, and thus it is a square matrix.
The above index theorem (38) applied to this $M$ gives rise to 
\begin{equation}
dim\ ker\ \slash{C}(\frac{1+\gamma_{5}}{2}) - dim\ ker\ \slash{C}(\frac{1- \gamma_{5}}{2})=0
\end{equation}
This shows that the possible zero modes of $\slash{C}$ appear in chiral pairs for {\cal any} background gauge field configuration. Namely, $Tr\gamma_{5}=0$ for the 
basis set defined by
\begin{equation}
\slash{C}\varphi_{n}=\lambda_{n}\varphi_{n}
\end{equation}
since $\varphi^{\dagger}_{n}\gamma_{5}\varphi_{n}=0$ for $\lambda_{n}\neq 0$. Here one may recall that for continuum operator $\slash{D}$ , we have the Atiyah-Singer index theorem,  $dim\ ker\ \slash{D}(\frac{1+\gamma_{5}}{2}) - dim\ ker\ \slash{D}(\frac{1- \gamma_{5}}{2})=\nu$,  or $Tr\gamma_{5}=\nu$ in a suitably regularized sense.   
The above index relation (40) is related to  a  statement of the species doubling 
for $\slash{C}$. In fact, a smooth continuum limit of the index relation (40)
 on the lattice for any background gauge field is consistent only with species doubling. It is important to recognize that the relation  $Tr \gamma_{5}=0$,
which appears to be almost self-evident in lattice theory, sometimes contains
more information than simply meaning that $Tr\gamma_{5}$ is a c-number.

For a square lattice one can explicitly show that the simplest lattice fermion action
\begin{equation}
S = \bar{\psi}i\slash{C}\psi
\end{equation}
is invariant under the transformation[15]
\begin{equation}
\psi^{\prime}= {\cal T}\psi,\ \bar{\psi}^{\prime}= \bar{\psi}{\cal T}^{-1}
\end{equation}
where ${\cal  T}$ stands for any one of the following 16 operators
\begin{equation}
1,\ T_{1}T_{2},\ T_{1}T_{3},\ T_{1}T_{4},\ T_{2}T_{3},\ T_{2}T_{4},\ T_{3}T_{4},\ T_{1}T_{2}T_{3}T_{4}
\end{equation}
and 
\begin{equation}
T_{1},\ T_{2},\ T_{3},\ T_{4},\ T_{1}T_{2}T_{3},\ T_{2}T_{3}T_{4},\ T_{3}T_{4}T_{1},\ T_{4}T_{1}T_{2}
\end{equation}
The operators  $T_{\mu}$  are  defined by 
\begin{equation}
T_{\mu}\equiv \gamma_{\mu}\gamma_{5}\exp {(i\pi x^{\mu}/a)}  
\end{equation}
and  satisfy the relation
\begin{equation}
T_{\mu}T_{\nu} + T_{\nu}T_{\mu}=2\delta_{\mu\nu}
\end{equation}
with  $T_{\mu}^{\dagger} = T_{\mu} = T^{-1}_{\mu}$ for anti-hermitian $\gamma_{\mu}$. 
We denote the 16 operators by ${\cal T}_{n}, \ \ n=0\sim 15$, in the following 
with ${\cal T}_{0}=1$.
By recalling that the operator $T_{\mu}$ adds the  momentum $\pi/a$ to the 
fermion momentum $k_{\mu}$, we cover the entire Brillouin zone in (25) by the operation (43) starting with the free fermion defined in
\begin{equation}
- \frac{\pi}{2a} \leq k_{\mu} \leq \frac{\pi}{2a}
\end{equation}
The operators in (44) commute with $\gamma_{5}$, whereas those in (45) anti-commute with $\gamma_{5}$ and thus change the sign of  chiral charge,  reproducing the species doublers with correct chiral charge assignment; $\sum_{n=0}^{15}
(-1)^{n}\gamma_{5} =0$. 

We now evaluate the index associated with the Jacobian factor defined in (37)
\begin{eqnarray}
- 2 Tr \gamma_{5}(1-\frac{a}{2}D)&=&- 2Tr\gamma_{5} + Tr \gamma_{5}(1+X\frac{1}{\sqrt{X^{\dagger}X}})\nonumber\\
&=&  Tr\gamma_{5} (X\frac{1}{\sqrt{X^{\dagger}X}})
\end{eqnarray}
by noting $Tr\gamma_{5}=0$ for our choice of the basis set. A local version of the index is given by ( here trace stands for Dirac and Yang-Mills indices )
\begin{equation}
 tr\gamma_{5} (X\frac{1}{\sqrt{X^{\dagger}X}}) 
\end{equation}
The behavior of the operator $X\frac{1}{\sqrt{X^{\dagger}X}}$ in the continuum limit is marginal ($X\frac{1}{\sqrt{X^{\dagger}X}}$ is a phase factor ), and to make the operation of taking a continuum limit of the above local index  better defined,  we supply an auxiliary regulator $h(x)$ as 
\begin{equation}
 tr \gamma_{5} [X\frac{1}{\sqrt{X^{\dagger}X}}h(\frac{\slash{C}^{2}}{M^{2}})] 
\end{equation}
The regulator $h(\slash{C}^{2}/M^{2})$ with $h(0)=1$, which is gauge invariant and ensures  the condition 
$Tr \gamma_{5}h(\slash{C}^{2}/M^{2})=0$ we are assuming, makes the manipulation at intermediate steps better defined but the final result is not influenced by $h(x)$, as will be seen later. We take an infinite volume limit $L=Na\rightarrow \infty$ in the expression (51) to accommodate the continuous momentum range in (25). 
We now change the basis to  plane wave basis defined in the momentum range in (48) combined with the transformation ${\cal T}_{n}$.

To make the analysis of the continuum limit more transparent, we also adopt the following two-step process: We first take the limit $a\rightarrow 0$ with 
\begin{equation}
\frac{2r}{a}, \ \ \frac{m_{0}}{a}
\end{equation}
kept fixed, and then later take the limit $\frac{2r}{a}, \  \frac{m_{0}}{a}\rightarrow \infty$. It is known that this two-step procedure  makes the perturbative  calculation in a continuum limit of the Wilson fermion easier[15]: In our context, this two-step procedure allows us to reduce the calculation of (50) to a calculation of a local index of continuum $\slash{D}$, a satisfactory aspect from a viewpoint of index theorem.

We thus obtain ( by keeping $\epsilon_{n}= (M_{n}/M)^{2}$ small not to disturb  physical contents )
\begin{eqnarray}
 && \lim_{a\rightarrow 0}tr \gamma_{5} [X\frac{1}{\sqrt{X^{\dagger}X}}h(\frac{\slash{C}^{2}}{M^{2}})]\nonumber\\
&=& \sum^{15}_{n=0}\lim_{a\rightarrow 0}tr\int_{-\frac{\pi}{2a}}^{\frac{\pi}{2a}}\frac{d^{4}k}{(2\pi)^{4}}e^{-ikx}{\cal T}^{-1}_{n}\gamma_{5}X\frac{1}{\sqrt{X^{\dagger}X}}h(\frac{\slash{C}^{2}}{M^{2}}){\cal T}_{n}e^{ikx}\nonumber\\
&=& \sum^{15}_{n=0}\lim_{L\rightarrow\infty}\lim_{a\rightarrow 0}tr\int_{-L}^{L}\frac{d^{4}k}{(2\pi)^{4}}e^{-ikx}\gamma_{5}(-1)^{n}X_{n}\frac{1}{\sqrt{X_{n}^{\dagger}X_{n}}}h(\frac{\slash{C}^{2}}{M^{2}})e^{ikx}\nonumber\\
&=& \sum^{15}_{n=0}\lim_{L\rightarrow\infty}tr\int_{-L}^{L}\frac{d^{4}k}{(2\pi)^{4}}e^{-ikx}\gamma_{5}(-1)^{n}(-i\slash{D}+M_{n})\frac{1}{\sqrt{\slash{D}^{2} + M_{n}^{2}}}h(\frac{\slash{D}^{2}}{M^{2}})e^{ikx}\nonumber\\
&=& \sum^{15}_{n=0}tr\int_{-\infty}^{\infty}\frac{d^{4}k}{(2\pi)^{4}}e^{-ikx}\gamma_{5}(-1)^{n}M_{n}\frac{1}{\sqrt{\slash{D}^{2} + M_{n}^{2}}}h(\frac{\slash{D}^{2}}{M^{2}})e^{ikx}
\end{eqnarray}
where we defined 
\begin{equation}
X_{n}\equiv {\cal T}^{-1}_{n}X{\cal T}_{n}= i\slash{C} + {\cal T}^{-1}_{n}B{\cal T}_{n}- \frac{m_{0}}{a}
\end{equation}
The second term in the right-hand side of $X_{n}$ gives rise to the 
mass term produced   by  the Wilson term in our continuum limit with any fixed 
$k_{\mu},  (-\frac{\pi}{2a}< k_{\mu}< \frac{\pi}{2a}),$ and $a\rightarrow 0$. We recall  that the momentum representation of $X$ for the vanishing gauge field is given by
\begin{equation}
\sum_{\mu}\gamma^{\mu}\frac{\sin ak_{\mu}}{a} + \frac{r}{a}\sum_{\mu}(1 -
\cos ak_{\mu}) - \frac{m_{0}}{a}
\end{equation}
The mass parameter $M_{n}$ thus stands for $M_{0}= - \frac{m_{0}}{a}$ and one 
of 
\begin{eqnarray}
&&\frac{2r}{a}-\frac{m_{0}}{a},\ \ (4,-1);\ \ \ 
\frac{4r}{a}-\frac{m_{0}}{a},\ \ (6,1)\nonumber\\
&&\frac{6r}{a}-\frac{m_{0}}{a},\ \ (4,-1);\ \ \ 
\frac{8r}{a}-\frac{m_{0}}{a},\ \ (1,1)
\end{eqnarray}
for $n=1\sim 15$. Here we specified ( multiplicity, chiral charge ) in the bracket for species doublers.
We also used the relation
\begin{eqnarray}
X e^{ikx}g(x) &\rightarrow& e^{ikx}(\slash{k}-i\slash{\partial}-\slash{A} + M_{0})g(x)\nonumber\\
&=&(-i \slash{D}+M_{0})(e^{ikx}g(x))
\end{eqnarray}
for $a\rightarrow 0$ with any fixed $k_{\mu}, (- \frac{\pi}{2a} < k_{\mu} <
\frac{\pi}{2a}$ ), and fixed $\frac{m_{0}}{a}$ for a sufficiently smooth function $g(x)$.
It is known that
\begin{equation}
M_{0}< 0, \ \ M_{n}> 0, n= 1\sim 15
\end{equation}
i.e., $2r> m_{0} > 0$, is required to have a single massless fermion pole in the propagator $1/D$ [2].
In the above expression (53), we  used  $\gamma_{5}\slash{D}+ \slash{D}\gamma_{5}=0$ and the cyclic property of trace operation. 

Eq.(53) thus becomes 
\begin{eqnarray}
 && - tr\int_{-\infty}^{\infty}\frac{d^{4}k}{(2\pi)^{4}}e^{-ikx}\gamma_{5}\frac{1}{\sqrt{\slash{D}^{2}/M_{0}^{2}+1 }}h(\epsilon_{0}(\slash{D}^{2}/M_{0}^{2}))e^{ikx}\nonumber\\
&&+ \sum^{15}_{n=1}tr\int_{-\infty}^{\infty}\frac{d^{4}k}{(2\pi)^{4}}e^{-ikx}\gamma_{5}(-1)^{n}\frac{1}{\sqrt{\slash{D}^{2}/M_{n}^{2}+1 }}h(\epsilon_{n}(\slash{D}^{2}/M_{n}^{2}))e^{ikx} 
\end{eqnarray}
by noting $M_{0}< 0$. All the terms in this expression have the standard form 
of a local index in (12) with
\begin{equation}
f(x) = \frac{1}{\sqrt{x + 1}}h(\epsilon_{n}x)
\end{equation}
which satisfies the condition (16). ( At this last stage, one may even take the  limit $\epsilon_{n}= (M_{n}/M)^{2}\rightarrow 0$ by using $h(0)=1$.)
We thus obtain in the limit $|M_{0}|$ and $ M_{n}\rightarrow \infty$
\begin{equation}
 (\sum_{n=1}^{15}(-1)^{n}-1)(\frac{g^{2}}{32\pi^{2}})tr \epsilon^{\mu\nu\alpha\beta}F_{\mu\nu}F_{\alpha\beta}
= -2 (\frac{g^{2}}{32\pi^{2}})tr \epsilon^{\mu\nu\alpha\beta}F_{\mu\nu}F_{\alpha\beta}
\end{equation}
which gives rise to the correct chiral anomaly and a local version of  the index theorem if one recalls (49): The Jacobian factor is given by $\ln J = -2i\int dx \alpha 
(\frac{g^{2}}{32\pi^{2}})tr \epsilon^{\mu\nu\alpha\beta}F_{\mu\nu}F_{\alpha\beta} $. 

From the present explicit calculation in (53), we learn that the Ginsparg-Wilson relation (1) by itself does not specify the 
coefficient of the anomaly factor uniquely: For example, $\ln J =0$ for 
$m_{0} < 0$, and the coefficient of the Jacobian becomes $6i\alpha$ instead of $-2i\alpha$ for $4 r > m_{0} > 2 r$. 
The condition of the absence of species doublers in $D$, $2r>m_{0}>0$, determines the anomaly uniquely. 
These properties are in accord  with an explicit weak coupling analysis in Ref.[16], where the coefficient of the anomaly is shown to be invariant under a wide range of continuous deformation of parameters provided certain inequalities are 
satisfied; their analysis is also consistent with our two-step procedure in taking a continuum limit.
  
It can be confirmed that the operator $D$ in (35) itself has the following property for
$a\rightarrow 0$ with any fixed $k_{\mu}, \ (-\frac{\pi}{2a}<k_{\mu}<\frac{\pi}{2a}),$
\begin{eqnarray}
D_{0}e^{ikx} &\rightarrow& -i\slash{D}e^{ikx}\nonumber\\
D_{n}e^{ikx} &\rightarrow& \frac{2}{a}e^{ikx}, \ \ \ n\neq 0
\end{eqnarray}
where $D_{n}\equiv {\cal T}_{n}^{-1}D{\cal T}_{n}$ with $2r> m_{0}=1 >0$. This property  is sufficient to prove a local version of the index theorem  (27)
 in Section 3.

Although we obtain an identical result at the end, the actual calculational
schemes in Section 3 and Section 4 are quite different. In Section 3 the basic operator is $\gamma_{5}D$, whereas the basic operator is the Wilson fermion
operator appearing in $aD$ in the present Section. From a viewpoint of index theorem, 
the calculational scheme  in Section 3 is more natural and much simpler. It should be noted that we deal with only finite quantities in both calculational schemes in Sections 3 and 4.      

\section{Discussion and conclusion}  
We studied the  implications of the recent development in lattice gauge theory on the formulation of chiral anomaly in continuum theory. For this purpose, we evaluated the chiral Jacobian in lattice theory in a continuum limit without referring to perturbative expansion. In lattice theory , we deal with well-defined finite quantities everywhere, and thus the identification of the Jacobian factor , in particular, the procedure from (10) to (11) in continuum theory is  better specified. Once one arrives at (11), one can perform a well-defined operation both in continuum and latttice theory. In fact, we formulated 
the index theorem on the lattice in an analogous manner in eq.(21), which leads to the master formula of anomaly in (27) on the basis of a few fundamental properties of  the lattice operator $D$. 

The conventional continuum regularization such as the Pauli-Villars regularization or dimensional regularization introduces an explicit chiral symmetry breaking into the Lagrangian, and at the same time the regularization eliminates the chiral Jacobian factor in path integral. In contrast, the lattice formulation [5]
 introduces the chiral Jacobian factor in a regularized theory, and thus naturally leads 
to the path integral formulation of chiral anomaly in continuum theory [8]
in a suitably defined continuum limit.
 This may suggest that the lattice regularization may 
eventually lead to a natural formulation of chiral symmetry even for chiral gauge theory. 

A major advantage of lattice theory lies in the fact that it allows a non-
perturbative numerical simulation of strongly coupled gauge theory. In this 
respect, the index theorem with a finite cut-off [4] may play an important role in such analyses. Both of fermion and gauge field configurations in our formulation of a continuum limit correspond to those of the conventional continuum theory. On the other hand, the field configurations appearing in the lattice formulation in Refs.[1][4] retain much of local lattice structures. For example, the 
fermion spectrum contains a non-linear dispersion relation. In this respect, the index theorem with a finite cut-off  may be regarded as a semi-local version of index theorem. The global Atiyah-Singer index 
theorem is based on the precise global topological properties of space-time
(such as $S^{4}$) and gauge fields. On the other hand, a local version of the index theorem, i.e., the chiral anomaly is defined independently of global topological structures of space-time and gauge fields. The index theorem with a finite cut-off in Ref.[4] may characterize the properties of space-time and gauge fields in-between these two limits known in continuum theory. It is quite exciting if one can confirm such a possibility by further analyses.\\ 
\\
{\bf Note added}\\
Recently, D.H. Adams  and also H. Suzuki[17] performed a (non-perturbative) evaluation of lattice chiral Jacobian for the overlap Dirac operator by using a calculational scheme different from ours. The final results agree with ours in 
Section 4 including the parameter dependence.

\end{document}